# Demonstration of 3 V Programmable Josephson Junction Arrays Using Non-Integer-Multiple Logic

Wenhui Cao[1], Erkun Yang[2], Jinjin Li[1], Guanhua She[1], Yuan Zhong[1], Qing Zhong[1], Da Xu[1], Xueshen Wang[1], Xiaolong Xu[1], Shijian Wang[1] and Jian Chen[1]

[1] Center for advanced measurement science, National Institute of Metrology, Beijing, China
[2] Xidian University, Xi'an, China

E-mail: jinjinli@nim.ac.cn and yangerkun@xidian.edu.cn

## Abstract

This article demonstrates a new kind of programmable logic for the representation of an integer that can be used for the programmable Josephson voltage standard. It can enable the numbers of junctions in most bits to be variable integer values, which is different from normal binary logic or ternary logic. Consequently, missing junctions due to superconducting short circuits can be tolerated under this logic. This logic can also have nearly the same segmentation efficiency as ternary logic. The completeness of the sequences using this logic is proven by the recursive method in mathematics in this paper. After that, a new algorithm for the representation of integers is presented according to the proven process, and an analysis of the number of fault-tolerant junctions for each bit is provided. Although the first and second bits are not tolerant to missing junctions, bits beyond these can tolerate one to hundreds of missing junctions. Due to the non-fixed multiples between the bits of the sequence, this logic is called non-integer-multiple logic. Finally, the design and fabrication of a 3 V programmable Josephson junction array using this logic are described, and the measurements and analysis of the characteristic parameters are presented.

## 1. Introduction

The output voltage of a programmable Josephson junction array (PJJA) satisfies the relationship $V = m \dfrac{f}{K_J}$, where $K_J$ is the Josephson constant, $m$ is the number of Josephson junctions (JJs), and $f$ is the operating frequency of the junction array [1–3]. According to the different methods of expressing integers $m$, the existing programmable arrays normally follow binary logic [4–7] or ternary logic [8–10]. For ternary logic PJJAs, one of the sub-arrays is usually divided into small sub-arrays in a threefold manner as follows: 1 ($3^0$), 3 ($3^1$), 9 ($3^2$), 27 ($3^3$), …, 6561 ($3^8$). The small sub-arrays are called the least significant bits (LSBs). Other large equal sub-arrays are called the most significant bits (MSBs) [11, 12].

LSBs and MSBs often have missing JJs owing to insulation defects between the wiring and base electrode or particles among the junction areas, resulting in superconducting shorts in one or both double-stacked barriers [6, 8, 12–17]. This failure is not critical, and a flat quantized step at a voltage corresponding to the remaining number of junctions in that sub-array can be measured. When junctions are missing on the LSBs, such as when a 9-junction bit becomes an 8-junction bit, the number relationship between the LSBs no longer follows the triple relationship, resulting in an incomplete sequence; that is, some numbers *m* cannot be expressed from the defective ternary sequence. Sometimes, even adjusting the frequency may not achieve a continuous voltage output of the programmable junction array, resulting in issues when outputting certain DC voltages or waveforms.

This article proposes a new type of programmable logic that can avoid this problem. Firstly, we demonstrate the relationship between the bits of the sequence under this logic. Secondly, we verify the completeness of the non-integer-multiple sequence (NIMS) from a mathematical perspective and introduce an algorithm for the representation of integers for this logic. Thirdly, we provide an analysis of the segmentation efficiency for NIMS. Finally, we describe the design and fabrication of a 3 V programmable Josephson junction array using this logic and present the measurements and analysis of the characteristic parameters.

## 2. Theory of non-integer-multiple logic

### 2.1 Definition of non-integer-multiple logic

The definition of non-integer-multiple logic is that, if the bits $a_n$ of a sequence obey the relationship:

$$0 < a_{n-1} < a_{n+1}/3 \leq a_n \leq 3a_{n-1} \quad (1)$$

where $n$ = 0, 1, 2, 3, ..., N, any integer *m* in the range $-\sum_{n=0}^{N} a_n \leq m \leq \sum_{n=0}^{N} a_n$ can be expressed as $m = \sum_{n=0}^{N} b_n a_n$, where $a_0 = 1$, $n$, N, and $a_n$ are integers, and $b_n$ is the sign of $a_n$, which can be chosen as −1, 0 or 1.

Each $a_n$ is close to, but not greater than, three times the preceding bit. Therefore, the relationship between bits is a non-integer-multiple value within the constraint range of three times, which we refer to as the NIMS. Due to the variable range of $a_n$ rather than a fixed value, it can be allowed a certain margin of error, which is given by $a_n - a_{n+1}/3$. Thus, except the first two bits, each bit can be designed to have several integer values and the sequence can remain complete. We will provide a verification in Section 3.

### 2.2 Mathematical proof

This section provides proof of the completeness of the aforementioned NIMS. In other words, when the value of a bit of a NIMS varies within the defined range, the sequence remains complete. In general, we set $a_0 = 1$, and the integer *m* represented by such a sequence is completely consistent with the expressed value. $a_0 = 2$ corresponds to the case of stacked junctions, where the difference between *m* and the expressed value is 1 or 0, and the coverage of all voltages can be achieved by fine-tuning the frequency. The first three bits satisfying the definition conditions (1) are listed in the following table (Table 1) for $a_0 = 1$.

**Table 1.** Possible integer combinations of NIMS with three bits starting from $a_0 = 1$

| $a_0$ | $a_1$ | $a_2$ |
|---|---|---|
| 1 | 2 | 3 |
| 1 | 2 | 4 |
| 1 | 2 | 5 |
| 1 | 3 | 4 |
| 1 | 3 | 5 |
| 1 | 3 | 6 |
| 1 | 3 | 7 |
| 1 | 3 | 8 |
| 1 | 3 | 9 |

Through manual calculations, we can determine that all these combinations are complete. Let us consider the sequence containing 1, 3, and 8 as an example. We will list the possible combinations of bits to express *m* as $m = \sum_{n=0}^{N} b_n a_n$.

**Table 2.** Sign of each bit to express *m*.

| *m* | Sign of 1 ($b_0$) | Sign of 3 ($b_1$) | Sign of 8 ($b_2$) |
|---|---|---|---|
| 0 | 0 | 0 | 0 |
| 1 | 1 | 0 | 0 |
| 2 | −1 | 1 | 0 |
| 3 | 0 | 1 | 0 |
| 4 | 1 | 1 | 0 |
| 5 | 0 | −1 | 1 |
| 6 | 1 | −1 | 1 |
| 7 | −1 | 0 | 1 |
| 8 | 0 | 0 | 1 |
| 9 | 1 | 0 | 1 |
| 10 | −1 | 1 | 1 |
| 11 | 0 | 1 | 1 |
| 12 | 1 | 1 | 1 |

Table 2 shows that all positive *m* can be expressed by a NIMS. Conversely, all negative *m* can be achieved by flipping the sign of $b_n$.

In the next step, we demonstrate the scalability of the NIMS. Based on the analysis described previously, we assume that $-\sum_{n=0}^{N} a_n \leq m \leq \sum_{n=0}^{N} a_n$ and *m* can be expressed as $m = \sum_{n=0}^{N} b_n a_n$; consequently, the proof must show that if *m'* is in the range $-\sum_{n=0}^{N+1} a_n \leq m' \leq \sum_{n=0}^{N+1} a_n$, then *m'* can also be expressed as $m' = \sum_{n=0}^{N+1} b_n a_n$. The subsequent analysis aims to verify the statement for positive integer numbers, whereas the case for negative numbers can be proven similarly by flipping the state of $b_n$.

Considering the first case, where $a_{N+1} \leq m' \leq \sum_{n=0}^{N+1} a_n$, we know that $m' - a_{N+1}$ is satisfied from the preconditions of $-\sum_{n=0}^{N} a_n \leq m' - a_{N+1} \leq \sum_{n=0}^{N} a_n$ and can be expressed as $m' - a_{N+1} = \sum_{n=0}^{N} b_n a_n$. Therefore, we can select $b_{N+1} = 1$ to express *m'* in terms of $a_n$ as $m' = a_{N+1} + \sum_{n=0}^{N} b_n a_n$, or equivalently as $m' = \sum_{n=0}^{N+1} b_n a_n$.

Considering the second case where $\sum_{n=0}^{N} a_n < m' < a_{N+1}$. When n=N+1, we get from the definition inequality (1), that $a_{N+1} \leq 3a_N$, so we have

$$0 < a_{N+1} - m' < 3a_N - \sum_{n=0}^{N} a_n \quad (2)$$

For the right side of this inequality, we can rewrite as:

$$3a_N - \sum_{n=0}^{N} a_n = 2a_N - \sum_{n=0}^{N-1} a_n \quad (3)$$

For n=N, we can also get form the definition inequality (1), that $a_N \leq 3a_{N-1}$, so equation (3) can be change to inequality as below:

$$3a_N - \sum_{n=0}^{N} a_n \leq a_N + 3a_{N-1} - \sum_{n=0}^{N-1} a_n \quad (4)$$

and $3a_{N-1} - \sum_{n=0}^{N-1} a_n \leq a_{N-1} + 3a_{N-2} - \sum_{n=0}^{N-2} a_n$

so,

$$3a_N - \sum_{n=0}^{N} a_n \leq a_N + a_{N-1} + 3a_{N-2} - \sum_{n=0}^{N-2} a_n$$

At last we can get:

$$3a_N - \sum_{n=0}^{N} a_n \leq \sum_{n=0}^{N} a_n + a_0 \quad (5)$$

So, inequality (2) can be rewrite as:

$0 < a_{N+1} - m' < \sum_{n=0}^{N} a_n + a_0$. As $a_0$ is the smallest bit, if $a_0$ is removed from the right side of the inequality, we have $0 < a_{N+1} - m' \leq \sum_{n=0}^{N} a_n$.

According to the assumption, $a_{n+1}-m'$ can be expressed as $a_{N+1} - m' = \sum_{n=0}^{N} b_n a_n$, yielding $m' = a_{N+1} - \sum_{n=0}^{N} b_n a_n$. If we let $b_n' = -b_n$ and $b_{N+1}' = 1$, then *m'* can be expressed as $m' = \sum_{n=0}^{N} b_n' a_n + b_{N+1}' a_{N+1} = \sum_{n=0}^{N+1} b_n' a_n$. Thus, the NIMS has been proven to be expandable and complete. Therefore, under the definition inequality (1), NIMS is a complete sequence.

## 2.3 Algorithm for integer representation using NIMS

The previously described proof provides a calculation algorithm for expressing *m'*. For $a_{N+1} \leq m' \leq \sum_{n=0}^{N+1} a_n$, *m'* can be obtained using $a_{N+1} - m$; for $\sum_{n=0}^{N} a_n < m' < a_{N+1}$, *m'* can be obtained using $a_{N+1} + m$. Thus, for $\sum_{n=0}^{N} a_n < m' \leq \sum_{n=0}^{N+1} a_n$, $a_{N+1}$ is selected as one of the bits to represent $m'$ and $b_{N+1}$ is set to 1. When $\sum_{n=0}^{N} a_n < m' < a_{N+1}$, the remainder of $m'$ will become negative, and we can denote the sign and compare the absolute value of the remainder with $\sum_{n=0}^{N} a_n$ (N = 0, 1, 2, …, N) to find the remaining bits for representation of $m'$, and so on. As $m$ is the remainder obtained after $a_{N+1}$ is selected for $m'$, $m$ is subsequently compared with the rest of $a_n$ (n = 0, 1, 2, 3, ..., N), and the previously described process is repeated until the remainder is zero.

From the above description, an algorithm for computer program to expressing $m$ is shown in Fig. 1, where $A_N = \sum_{n=0}^{N} a_n$, $C_n = \sum_{n=0}^{n} a_n + a_0$, and $s$ denotes the sign of $m$ and each bit during calculation.

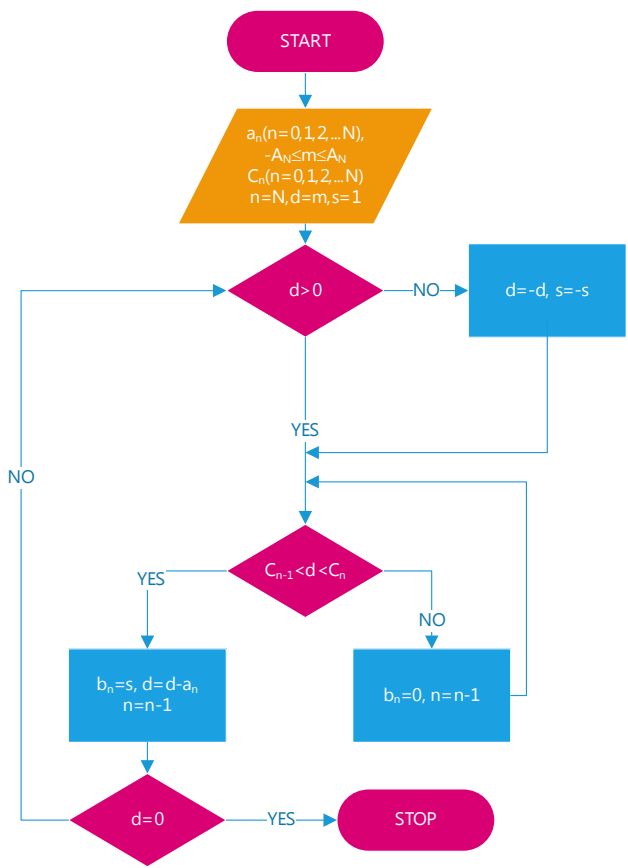

Fig. 1 Algorithm for expressing integer $m$ using the NIMS.

Because the ternary or binary bits also satisfy the definition of the NIMS inequality $0 < a_{n-1} < a_{n+1}/3 \leq a_n \leq 3a_{n-1}$, this algorithm can be used not only to identify all non-integer multiples bits that match $m$, but also for ternary or binary sequences to express $m$. In a future study on mathematical programming, we will compare this algorithm with the balanced ternary or binary calculation algorithms.

## 3. Design of PJJA using NIMS

Table 3 presents an ordinary binary sequence, a ternary sequence, and two NIMSs (NIMS 1 and NIMS 2). For a PJJA, one or two of the larger sub-arrays should be divided into smaller bits to form a binary or ternary sequence. The smallest bit represents the voltage resolution of the programmable junction array. The ternary sequence requires fewer small bits to achieve high resolution than the binary sequence. However, for a NIMS, nearly the same number of small bits is required to achieve a high resolution. If we define $a_{n+1}/a_n$ as the segmentation efficiency, the NIMS can achieve an efficiency comparable to that of a standard ternary sequence.

**Table 3.** Examples of binary sequence, ternary sequence, and two NIMSs for PJJA.

| Method | Binary | Ternary | NIMS 1 | Fault-tolerant numbers | NIMS 2 | Fault-tolerant numbers |
|---|---|---|---|---|---|---|
| $a_0$ | 1 | 1 | 1 | 0 | 2 | 0 |
| $a_1$ | 2 | 3 | 2 | 0 | 6 | 0 |
| $a_2$ | 4 | 9 | 6 | 1 | 18 | 2 |
| $a_3$ | 8 | 27 | 14 | 1 | 48 | 4 |
| $a_4$ | 16 | 81 | 39 | 1 | 132 | 6 |
| $a_5$ | 32 | 243 | 114 | 2 | 378 | 6 |
| $a_6$ | 64 | 729 | 336 | 4 | 1116 | 12 |
| $a_7$ | 128 | 2187 | 996 | 6 | 3312 | 20 |
| $a_8$ | 256 | 6561 | 2970 | >100 | 8800 | >100 |
| $a_9$ | 512 | 8000 | 8000 | >100 | 8800 | >100 |
| $a_{10}$ | 1024 | 8000 | 8000 | >100 | 8800 | >100 |
| $a_{11}$ | 2048 | 8000 | 8000 | >100 | 8800 | >100 |
| $a_{12}$ | 4096 | 8000 | 8000 | >100 | 8800 | >100 |
| $a_{13}$ | 8000 | 8000 | 8000 | >100 | 8800 | >100 |

As discussed in Section 4, the number of missing junctions allowed in bit $a_n$ is given by $a_n - a_{n+1}/3$. The fault tolerance proportion for $a_n$ is given by $(a_n - a_{n+1}/3)/a_n = 1 - a_{n+1}/3a_n$. Thus, if $a_{n+1}$ is significantly less than $3a_n$, a large proportion of $a_n$ can be a tolerant value and the segmentation efficiency will become low. However, as $a_{n+1}$ approaches $3a_n$, the fault proportion of $a_n$ decreases while the segmentation efficiency increases. If $a_{n+1}$ exceeds $3a_n$, the completeness of the ternary bits is compromised.

To ensure that the desired value of $m$ can be expressed by a binary or ternary sequence, strict consistency is required. For the NIMSs suggested in Table 3, at least one missing junction can be tolerated starting from the third bit without affecting the accuracy of the $m$ expression. For larger junction bits, more missing junctions caused by superconducting short circuits are tolerated. For bits with thousands of junctions, more than 100 missing junctions are allowed. However, hundreds of junctions are unlikely to be lost during fabrication. Typically, only a few junctions are lost, making the fault-tolerant quantities sufficient for most bits. For bits $a_0$ and $a_1$, which contain only a few junctions, the probability of having missing junctions is negligible.

For Josephson junction arrays with stack structures, the initial value of bit $a_0$ is 2 or 3. The expression for $m$ becomes

$$m = \sum_{n=0}^{N} b_n a_n + \beta$$

, where $\beta$ is an integer and $-a_0 < \beta < a_0$. Therefore, a difference in one or two junctions may exist between $m$ and the sum of the selected bits. By slightly tuning the frequency, a continuous voltage output can be achieved.

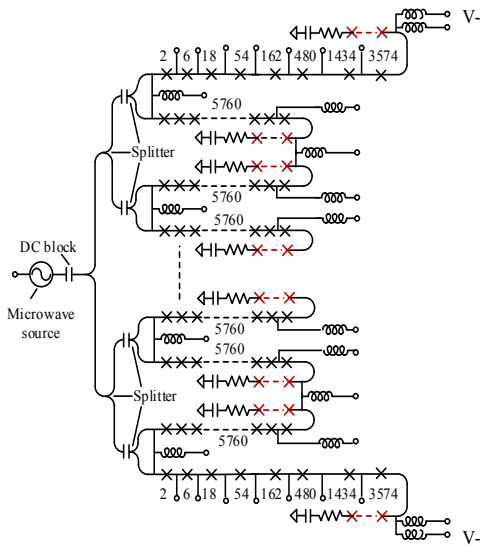

Fig. 2 Design of a 3 V NIMS programmable Josephson junction array.

Fig. 2 depicts the NIMS design of a 3 V programmable Josephson junction array using stacked Josephson junctions. Thus, in this design, $a_0$ is 2. The array is divided into 16 branches by four stages of Wilkinson microwave power dividers. The adjacent branches are connected by a resistance shunted with an inductance low-pass filter, and all the effective junctions are connected in series. Capacitances are present at the beginning and the terminal of every branch to ensure that the whole series array is floating to ground. The bilateral two large sub-arrays are divided into LSBs of NIMS. The fault tolerant number of Josephson junctions for each bit is analyzed in Table 5. It is difficult for bits with fewer than 100 Josephson junctions to have missing junctions. Thus, the starting several bits are designed to have no fault tolerant numbers. Due to this symmetrical design, two symmetrical parts can be subtracted to output a very small voltage to verify the flatness of the voltage step of the overall array.

### 4. Fabrication of PJJA

In a high-vacuum chamber, $Nb/(Nb_xSi_{1-x}/Nb)_2$ layers are sputtered in situ. The Nb base and top layers are deposited at a sputtering power of 500 W and a deposition rate of 0.75 nm/s. The $Nb_xSi_{1-x}$ film is co-sputtered. The position of the Nb and Si target are reorganized compared with the previous work. The sputtering powers for Nb and Si at 36 W and 230 W, respectively, resulting in a deposition rate of approximately 0.15 nm/s for this work. The barrier and middle Nb layers have thicknesses of 33 nm and 75 nm, respectively.

The top Nb layer, alternating Nb middle layer, and $Nb_xSi_{1-x}$ barrier layer are etched using reactive ion etching (RIE) plus inductively coupled plasma. $SF_6$ is used as an etching gas, while $C_4F_8$ serves as a passivating gas to protect the junction profile and maintain sharp, clean side walls for uniformity in the stacked junctions. $CF_4$ and $O_2$ are used as etching gas for Nb base electrode.

$SiO_2$ is deposited as an insulating layer using a plasma-enhanced chemical vapor deposition device. For etching $SiO_2$, $CHF_3$ and $O_2$ are utilized. Prior to depositing a 500-nm-thick Nb wiring electrode, the connection area undergoes DC-biased RF cleaning with a dc voltage of 283V for 5 mins.

Improvements in wire layer etching are achieved by optimizing the process using ICP and RIE with $CF_4$ and $O_2$ mixture gases, compared to previous work.[]

Subsequently, the sample surface is subjected to a 90-second RF cleaning of the oxide layer at 50W RF power and approximately 550V DC bias voltage, followed by dc-magnetron sputtering deposition of a 180 nm PdAu film. The sheet resistance is 1.9 ohms per square.

### 5. Measurement

**Table 5.** AC characteristics for each bit of a NIMS circuit. There are two junctions in each stack.

| Parameters | |
|---|---|
| Number of junctions | 92098 |
| Junction length (μm) | 11 |
| Junction width (μm) | 4 |
| Current density (kA/cm$^2$) | 18 |
| Critical current (mA) | 8 |
| Normal resistance (mΩ) | 4.2 |

The parameters of the circuit are measured at 4.2 K as shown in Table 4. The margins and junction numbers for each sub array with the circuit biased at 18.01 GHz are shown in Table 5. The positive and negative steps for each bit are larger than 1 mA.

**Table 5.** AC characteristics for each bit of a NIMS circuit. There are two junctions in each stack.

| NIMS | Junction number | Positive step width (mA) | Zero step width (mA) | Negative step width (mA) | Fault-tolerant numbers |
|---|---|---|---|---|---|
| $a_0$ | 2 | 3.96 | 9.18 | 4.01 | 0 |
| $a_1$ | 6 | 3.08 | 10.56 | 3.08 | 0 |
| $a_2$ | 18 | 3.74 | 7.64 | 3.79 | 0 |
| $a_3$ | 54 | 3.79 | 7.15 | 3.79 | 0 |
| $a_4$ | 162 | 2.86 | 7.97 | 2.47 | 2 |
| $a_5$ | 480 | 1.87 | 6.27 | 1.98 | 2 |
| $a_6$ | 1434 | 1.21 | 5.61 | 1.59 | 242 |
| $a_7$ | 3574 | 2.03 | 7.97 | 2.09 | >100 |
| $a_8$ | 5759 | 2.03 | 6.87 | 1.98 | >100 |
| $a_9$ | 5760 | 2.14 | 7.53 | 1.76 | >100 |
| $a_{10}$ | 5760 | 1.76 | 7.42 | 1.81 | >100 |
| $a_{11}$ | 5760 | 1.87 | 6.71 | 1.92 | >100 |
| $a_{12}$ | 5760 | 1.98 | 7.09 | 1.92 | >100 |
| $a_{13}$ | 5760 | 1.65 | 6.54 | 1.54 | >100 |
| $a_{14}$ | 5759 | 1.38 | 4.56 | 1.37 | >100 |
| $a_{15}$ | 5760 | 1.65 | 4.56 | 4.56 | >100 |
| $a_{16}$ | 5760 | 1.81 | 6.16 | 1.81 | >100 |
| $a_{17}$ | 5760 | 1.92 | 6.27 | 1.92 | >100 |
| $a_{18}$ | 5760 | 1.98 | 6.10 | 1.98 | >100 |
| $a_{19}$ | 5760 | 1.92 | 6.93 | 1.98 | >100 |
| $a_{20}$ | 5760 | 2.20 | 7.53 | 2.20 | >100 |
| $a_{21}$ | 5760 | 2.42 | 6.87 | 2.47 | >100 |
| $a_{22}$ | 5730 | 1.21 | 5.88 | 1.26 | >100 |

The total array has 92,098 Josephson junctions and can output a voltage of approximately 3.2 V. Combining the algorithm mentioned in section 4, we can achieve output voltages ranging from 2.5 mV to 3.2 V using this NIMS circuit. Fig. 3 shows the DC and AC characteristics of the whole array.

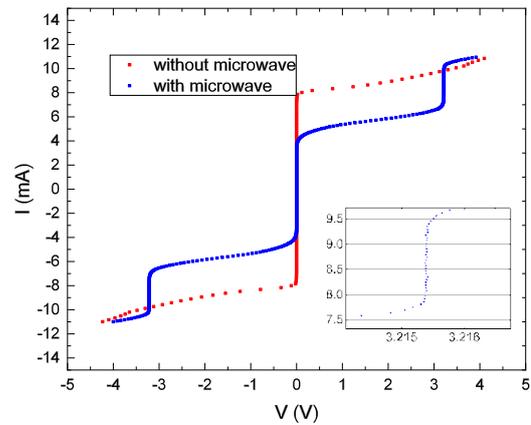

.Fig. 3 DC and AC I-V characteristics of the whole 3 V PJJA.

## 6. Conclusion

In this study, we propose a new kind of programmable logic for a programmable Josephson voltage standard and proved its completeness mathematically. Additionally, we developed an algorithm for expressing the desired number *m* using NIMSs, which can also be utilized for binary or ternary sequences. Compared with common binary and ternary sequences, the NIMS can accommodate missing junctions caused by superconducting short circuits and can achieve a segmentation efficiency comparable to that of a standard ternary sequence. Finally, a 3 V PJJA with NIMS was designed, fabricated, and measured. The characteristics of the whole array and each bit were analysed. The total array had 92,098 Josephson junctions and could output a voltage of approximately 3.2 V.